\documentclass[aps,prl,twocolumn,groupedaddress,nofootinbib,preprintnumbers]{revtex4}
\usepackage{graphicx}
\usepackage{dcolumn}
\usepackage{bm}
\usepackage{latexsym}
\usepackage{amsfonts}
\usepackage{amssymb}
\usepackage{amsmath}
\usepackage{verbatim}

\begin{document}


\title{Massive Graviton on a Spatial Condensate}

\author{Chunshan Lin}
\affiliation{Kavli Institute for the Physics and Mathematics of the Universe (WPI), Todai Institutes for Advanced Study, University of Tokyo, 5-1-5 Kashiwanoha, Kashiwa, Chiba 277-8583, Japan}


\begin{abstract}
In this paper, we consider the spatial gauge symmetries spontaneously break down in GR, and  graviton becomes massive on this spatial condensate background. Such model can be considered as a simplest example of massive gravity.  We then apply our massive gravity theory to inflation, the graviton mass removes the IR divergence of the inflationary loop diagram.
\end{abstract}

\maketitle

{\bf Introduction} In gauge field theory, the Higgs mechnism spontenously breaks the gauge symmetry, gives the gauge field a mass. Whether such mechanism can be applied to gravity and get a self-consistent massive spin-2 field theory is a basic question in the classical field theory. After the pioneering attampt at 1939 \cite{Fierz1939},  this direction has been attracting  a great deal of interest, but its consistency
has been a challenging problem for several decades.



One of the most profound problems of the massive gravity is the ghosty sixth mode in the gravity sector, which was found by Bolware and Deser in 1972 \cite{bdghost}. The BD ghost generally appears at the nonlinear massive gravity theory, where those nonlinear terms in the action were introduced to heal the discountinuity problem of the Fierz-Pauli theory  \cite{vdvz1}\cite{vdvz2}\cite{Vainshtein}. Because of the BD ghost, the Hamiltonian of the system is unbounded from below, which spoils the stability of our theory. 

An important breakthrough on the way of conquering the BD ghost was in 2002 \cite{ArkaniHamed:2002sp}. As pointed out by the authors of \cite{ArkaniHamed:2002sp}, by adopting the effective field theory at the decoupling limit, in principle we can eliminate the BD ghost by the construction of our massive gravity theory. Indeed, such type of theory was achieved in 2010, which now is dubbed as dRGT gravity \cite{deRham:2010kj}. 

However, the following up cosmological perturbations analysis revealed a new ghost instability among the rest five degrees of freedom \cite{Gumrukcuoglu:2011zh}\cite{DeFelice:2012mx}\cite{DeFelice:2013awa}\cite{DeFelice:2013bxa}\cite{Khosravi:2013axa}. On the other hand, this theory may also suffer from the acausality problem \cite{Deser:2012qx}\cite{Deser:2013eua}.

In this short notes, we propose an alternative construction of massive gravity. The idea is actually quite simple. We consider a spontaneous spatial symmetry breaking in GR. Such spatial gauge symmetry broken gives rise to 3 Goldstone excitations that were ``eaten" by graviton in the unitary gauge. The graviton gets mass and becomes a massive spin-2 particle, with 5 polarizations on the spectrum. The stablity of this theory is also carefully checked in this paper. This theory actually could be categorized as a Lorentz violation massive gravity theory. See\cite{Dubovsky2004}\cite{Rubakov:2008nh} for the early studies in this topic, and see \cite{deRham:2014zqa} for a recent review on massive gravity.


As an example of the application, we apply our massive gravity to early universe. It is known that Inflationary paradigm \cite{inflation} has become a very convincing scenario of the early universe. The quantum fluctuation during inflation seeds the large scale structure and CMB anisotropies nowadays. However, the power spectra of the primordial perturbation suffers from the infrared (IR)  divergence and ultraviolet (UV) divergence, if we take into account the contributions from the loop correction. These divergences were firstly noticed in the early work\cite{earlyloop1}\cite{earlyloop2}\cite{earlyloop3}, and  has been bothering the theorist for couple of decades (see the recent reviews \cite{Seery:2010kh}\cite{Tanaka:2013caa} and the references therein).


In this short notes, we focus on the IR divergence.  It is known that the scale invariant spectrum in the de-sitter space time leads to the logarithmically divergence in the IR.  
However, in the case of massive gravity, thanks to the graviton mass, the inflationary loop diagram converges at IR side.  

{\bf Spatial Condensation}
Firstly, let's write down such a simple action with Einstein-Hilbert term and 3 canonical massless scalar fields,
\begin{eqnarray}
S=M_p^2\int \sqrt{-g}\left(\frac{\mathcal{R}}{2}-\frac{1}{2}m^2g^{\mu\nu}\partial_{\mu}\phi^a\partial_{\nu}\phi^b\delta_{ab}-\Lambda\right),
\end{eqnarray}
where $\Lambda$ is the bare cosmological constant and $a,b=1,2,3$. The background solution spontaneously breaks the Lorentz invariance in terms of two different patterns. One is by spontaneously generating a preferred time direction, for example, the effective field theory of inflation\cite{Cheung:2007st}, and ghost condensation \cite{ArkaniHamed:2003uy}, where%
\begin{eqnarray}
\langle\phi^a\rangle=f(t)~,
\end{eqnarray}
and $f(t)$ is some function of time. In this case the graviton is still massless and thus it isn't the main interest of this paper. The second pattern spontaneously generates a preferred spatial frame,
\begin{eqnarray}\label{spcond}
\langle\phi^a\rangle=x^a~,
\end{eqnarray}
which gives us a spatial condensation scenario (see \cite{Endlich:2012pz} for a similar idea and its application in inflation). Please notice that at the l.h.s of above eq. (\ref{spcond}), the up index `a' is the internal index of scalar fields and $\phi^a$ remain invariant under the general coordinate transformation. However, at the r.h.s of equation, `a' is the space time index,  under the general coordinate transformation it changes as follows,
\begin{eqnarray}
x^a\to x^{a}+\xi^a~.
\end{eqnarray}
In order to maintain the eq.(\ref{spcond}) under the coordinate transformation, we introduce a Goldstone excitation $\pi^a$, which  transforms in the opposite way, 
\begin{eqnarray}
\phi^a=x^a+\pi^a~,~~~~\pi^a\to\pi^{a}-\xi^a.
\end{eqnarray}
The Goldstone excitations $\pi^a$ non linearly realize the diffeomorphisms and they describe the perturbations of 3 scalars. 

Our Goldstone excitations of such spatial condensation are actually a vector field, which can be decomposed into 3 independent components: one longitudinal mode and two transverse modes,
\begin{eqnarray}
\pi^a=\delta^{ab}(\partial_b\varphi+A_b)~.
\end{eqnarray}

 In the unitary gauge, we can see  those Goldstones are ``eaten" by the massless spin-2 field. After that, massless spin-2 particle gets weight and become massive, with 5 degrees of freedom on spectrum. In order to see how does this happen explicitly, let's do our honest perturbation calculations on the FRW background. Under the FRW ansatz, the metric reads
\begin{eqnarray}
ds^2=-N^2dt^2+a(t)^2dx^2~.
\end{eqnarray}
By taking the variation of the action with respect to the lapse and scale factor, we get the following two background Einstein equations,
\begin{eqnarray}\label{bgeom}
3H^2=\frac{3m^2}{2a^2}+\Lambda~,~~~~~~~~~\frac{\dot{H}}{N}=-\frac{m^2}{2a^2}~.
\end{eqnarray}
Then we perturb the space-time metric and define the metric perturbations by 
\begin{eqnarray}
g_{00}&=&-N^2(t)[1+2\phi]~,\\
g_{0i}&=&N(t)a(t)(S_i+\partial_i\beta)~,\\
g_{ij}&=&a^2(t)[\delta_{ij}+2\psi\delta_{ij}+(\partial_i\partial_j-\frac{1}{3}\partial^2)E\nonumber\\
&&~~~~~~~~~~~~+\frac{1}{2}(\partial_iF_j+\partial_jF_i)+\gamma_{ij}]~.
\end{eqnarray}
where
\begin{eqnarray}
\partial_iS^i=\partial_iF^i=\gamma^i_i=\partial_i\gamma^{ij}=0~.
\end{eqnarray}
Noting that the vector field defined by 
\begin{eqnarray}
Z^i\equiv\frac{1}{2}\delta^{ij}(\partial_j E+F_j)
\end{eqnarray}
transforms as 
\begin{eqnarray}
Z^i\to Z^i+\xi^i~.
\end{eqnarray}
Thus the combination $(Z^i+\pi^i)$ is a gauge invariant quantity. In the unitary gauge, $Z^i$ eats $\pi^i$, and survives in the linear perturbation theory. It is constrast to the general relativity, where $E$ and $F_i$ both are non-dynamical and we can just simply integrate them out. 

{\bf Scalar Perturbation} Now let's expand the action upto quadratic order in the unitary gauge, where $\phi^a=x^a$. For the scalar sector, we found that $\phi$, $\beta$ and $\psi$ are non-dynamical. After integrating out those non-dynamical modes, the quadratic action for the scalar perturbation reads
\begin{eqnarray}
\mathcal{L}_{s}\supset M_p^2\int dtd^3k\left( \frac{k^4m^2a^3N}{8k^2+12m^2}\frac{\dot{E}^2}{N^2}-\frac{k^2m^2(k^2+2m^2)aN}{8k^2+12m^2}E^2\right).\nonumber\\
\end{eqnarray}
Noted that background equations eq.(\ref{bgeom}) are used to get the above results.    As we expected, after eating the longitudinal mode of our Goldstone, the scalar metric perturbation $E$ survives and becomes a dynamical degree, propogates on the FRW space time background. By looking at the coefficient of the kinetic term, we can see it is always positive, as long as $m^2$ is positive. Thus our scalar mode is free from the ghost instability. 

 The canonical normalized scalar perturbation is defined by
\begin{eqnarray}
\mathcal{E}\equiv\frac{k^2M_pm\cdot E}{\sqrt{4k^2+6m^2}}~,
\end{eqnarray}
where $m$ is demanded to be positive. In terms of this canonical variable, the quadratic action for scalar perturbation can be rewritten as 
\begin{eqnarray}
\mathcal{L}_{s}\supset\frac{1}{2}\int dtd^3kNa^3\left(\frac{\dot{\mathcal{E}}^2}{N^2}-\omega_s^2\mathcal{E}^2\right)~,
\end{eqnarray}
where 
\begin{eqnarray}\label{sdisp}
\omega_s^2\equiv \frac{k^2}{a^2}+\frac{2m^2}{a^2}~.
\end{eqnarray}
From this dispersion realtion, we can see the sound speed of scalar mode is unity, and there is a mass gap on the scalar sepctrum. 

{\bf Vector Perturbation}
Now let's turn to the vector perturbation. We find the vector perturbation $S_i$ is non-dynamical and we can simply integrate it out. After that, the quadratic action of vector perturbation reads,
\begin{eqnarray}
\mathcal{L}_v\supset M_p^2\int dt d^3k\left(\frac{k^2m^2a^3N}{8k^2+16m^2}\frac{\dot{F}_i\dot{F}^i}{N^2}-\frac{k^2m^2aN}{8}F_iF^i\right).\nonumber\\
\end{eqnarray}
Similar to the scalar perturbation, in the unitary gauge, vector perturbation $F_i$ eats the transverse mode of our Goldstone, becomes a dynamical degree and propogates on the FRW space time background. 
By looking at the coefficient of kinetic term, our vector perturbation is also ghost free when $m^2$ is greater than zero. 

Then we canonical normalized the action by defining  such canonical variable, 
\begin{eqnarray}
\mathcal{F}_i\equiv\frac{kM_pm\cdot F_i}{2\sqrt{k^2+2m^2}},
\end{eqnarray}
and the quadratic action can be rewritten in terms of canonical variable as follows,
\begin{eqnarray}
\mathcal{L}_{v}\supset\frac{1}{2}\int dtd^3kNa^3\left(\frac{\dot{\mathcal{F}}_i\dot{\mathcal{F}}^i}{N^2}-\omega_v^2\mathcal{F}_i\mathcal{F}^i\right)~,
\end{eqnarray}
where 
\begin{eqnarray}\label{vdisp}
\omega_v^2\equiv \frac{k^2}{a^2}+\frac{2m^2}{a^2}~.
\end{eqnarray}
Due to the SO(3) symmetry of our scalar fields' configuration, the dispersion relation of vector mode is exactly the same as the one of scalar mode. 

{\bf Tensor Perturbation}
Now let's look at the final sector of our linear metric perturbation. After using the background equations, the quadratic action of our tensor modes reads,
\begin{eqnarray}
\mathcal{L}_T\supset M_p^2\int dtd^3k\left[\frac{a^3}{4N}\dot{\gamma}_{ij}\dot{\gamma}^{ij}-\frac{(k^2+2m^2)aN}{4}\gamma_{ij}\gamma^{ij}\right].\nonumber\\
\end{eqnarray}
Again, we do the canonical normalization,
\begin{eqnarray}
\tilde{\gamma}_{ij}\equiv\frac{M_p}{2}\gamma_{ij},
\end{eqnarray}
and the action can be rewritten as 
\begin{eqnarray}\label{tensoract}
\mathcal{L}_{T}\supset\frac{1}{2}\int dtd^3kNa^3\left(\frac{\dot{\tilde{\gamma}}_{ij}\dot{\tilde{\gamma}}^{ij}}{N^2}-\omega_T^2\tilde{\gamma}_{ij}\tilde{\gamma}^{ij}\right)~,
\end{eqnarray}
where 
\begin{eqnarray}\label{tensordisp}
\omega_T^2\equiv\frac{k^2}{a^2}+\frac{2m^2}{a^2}.
\end{eqnarray}
Surprisingly! The dispersion relation of our tensor mode is exactly the same as the one of scalar mode and vector mode.  On the other hand, our tensor mode receives a mass correction on the dispersion realtion, which is contrast to the general relativity. 


{\bf vDVZ discontinuity}
%
%
In the early and famous work of Fierz and Pauli \cite{Fierz1939}, the simplest linear extension to GR suffer from the vDVZ discontinuity, which the theory can not reduce to GR at the massless limit $m\to0$ \cite{vdvz1}\cite{vdvz2}\cite{Vainshtein}. One way to understand the origin of vDVZ discontinuity is to look at the decoupling limit, the scalar sector still couples tensor sector under such limit. 


In our spatial condensation scenario, such nontrivial coupling at the decoupling limit is absent, thus our theory can smoothly reduce to GR at the massless limit. At the massless limit, the effective action can be written in terms of a massless graviton and scalar mode of massive graviton,
\begin{eqnarray}
\mathcal{L}_{DL}&\supset& M_p^2\int\frac{1}{4}h^{\mu\nu}\mathcal{E}^{\alpha\beta}_{\mu\nu}h_{\alpha\beta}+\nonumber\\
&&m^2\left(-\frac{1}{2}k^2\dot{\varphi}^2+hk^2\varphi+hk^4\varphi^2+h^2k^2\varphi+...\right).\nonumber\\
\end{eqnarray}
Where $h_{\mu\nu}\equiv g_{\mu\nu}-\eta_{\mu\nu}$ and the indices are omitted for the simplicity of handwriting. The canonical normalized scalar and tensor modes are
\begin{eqnarray}
h^c\equiv M_p h~,~~~\varphi^c\equiv M_pmk\varphi~.
\end{eqnarray}
In terms of canonical variable, the linear coupling term between scalar and tensor is 
\begin{eqnarray}
mh^c\varphi^c\to0, 
\end{eqnarray}
it disappears at massless limit. The non-linear coupling terms between scalar and tensor are
\begin{eqnarray}
\frac{k}{M_p}kh^c\varphi^{c2}\to0~,~~\frac{mk}{M_p}h^{c2}\varphi^c\to0~,
\end{eqnarray}
which strongly suppressed by the factor of $k/M_p$ thus it can be neglected. One can easily check that the higher order coupling terms are also strongly suppressed by such factor.  Thus, we conclude that at the massless limit $m\to0$, our spatial condensate scalars decouple from gravity and we recover GR\footnote{A more convincing proof will be given in our upcoming work\cite{KLS}.}.


\begin{figure}
\begin{center}
\includegraphics[width=0.2\textwidth]{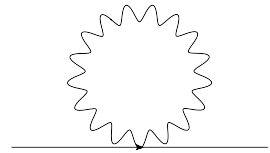}
\end{center}
\caption{one loop diagram, which corresponds to the scalar-tensor interaction.}
\label{loop}
\end{figure}

{\bf IR safe inflation}   
Since our spatial condensation is just free scalar theory,  the absence of higher order Goldstone interactions  implies that our effective field theory approach is valid up to the energy scale where the quantum gravity effect becomes important, say, Planck scale. 

As an example, we apply our massive gravity to early universe, see how does graviton mass remove the IR divergence.  During the inflation epoch, the action can be written as 
\begin{eqnarray}
S&=&\int \sqrt{-g}\left(\frac{M_p^2}{2}\mathcal{R}-M_p^2m^2\frac{1}{2}g^{\mu\nu}\partial_{\mu}\phi^a\partial_{\nu}\phi^b\delta_{ab}\right.\nonumber\\
&&~~~~~~~~~~~~~~\left. -\frac{1}{2}g^{\mu\nu}\partial_{\mu}\sigma\partial_{\nu}\sigma-V(\sigma)\right),
\end{eqnarray}
where $\sigma$ is the inflaton scalar. We take the one graviton loop diagram depicted in Fig.\ref{loop} as an example. This diagram is particularly important because if the inflaton is a free scalar field, such diagram makes the leading  contribution to the non-linear correction of the primordial spectrum\footnote{The authors of the paper\cite{Xue:2012wi} pointed out that such diagram is exactly canceled by another two-vertex loop diagram if $\dot{\epsilon}=0$, where $\epsilon$ is the slow roll parameter. However, $\epsilon$ is not always a constant for the most of inflationary models.}. The graviton interaction vertex corresponding to Fig.\ref{loop} is 
\begin{eqnarray}
\mathcal{H}_I\supset \gamma_{ij}^2\left(\partial_k\delta\sigma\right)^2~.
\end{eqnarray}
where $\delta\sigma$ is the inflaton scalar's perturbation, and $\gamma_{ij}$ is the tensor perturbations. We quantize the tensor mode as:
\begin{eqnarray}
\gamma_{ij}(x)=\sum_{s=\pm}\int d^3k\left[a(\bf{k})e_{ij}(k,s)\gamma_{k,s}e^{i\bf{k}\cdot\bf{x}}+h.c.\right],
\end{eqnarray}
where $a(k)$ is the annihilation operator, the subscript $s$ is the helicity, and $e_{ij}(k,s)$ is the transverse and traceless polarisation tensor which can be normalized as
\begin{eqnarray}
e_{ij}(k,s)e^{ij}(k,s')=\delta_{ss'}~.
\end{eqnarray}
The mode function in the de-sitter space time is easy to obtain. We assume that the fluctuation is generated at the deep sub-horizon scale, and the vacuum is the standard Bounch-David vacuum.  The quadratic action of tensor perturbations, i.e. eq.(\ref{tensoract})(\ref{tensordisp}), implies the mode function of the tensor perturbation takes such form,
\begin{eqnarray}
\gamma_{\pm,k}=\frac{H}{(2\pi)^{3/2}\sqrt{\tilde{k}^3}}(1+i\tilde{k}\eta)e^{-i\tilde{k}\eta}~,
\end{eqnarray}
where 
\begin{eqnarray}
\tilde{k}\equiv\sqrt{k^2+2m^2}~,
\end{eqnarray}
and the power sepcturm reads
\begin{eqnarray}
P_{GW}(\tilde{k})=\frac{2H^2}{(2\pi)^3\tilde{k}^3}\left[1+\mathcal{O}(\tilde{k}^2\eta^2)\right].
\end{eqnarray}

 We then calculate the one graviton loop in Fig \ref{loop}. 
Using in-in formalism, we find that one graviton loop diagram depicted in Fig. \ref{loop} obtained from the contraction between the two $\gamma$s, 
\begin{eqnarray}\label{nodiverg}
\langle\zeta(x)\zeta(x)\rangle_{1loop}&\propto&\int_0^{a(t)H(t)} d^3k P_{GW}(\tilde{k})\nonumber\\
&\propto& Ht+\log \left(H/m\right),
\end{eqnarray}
In the case of GR, such integral is divergent for sure. However, thanks to the graviton mass, the above loop integral is convergent at the IR side.  Noted that we only take into account the super horizon modes, thus the integral up bound can be chosen as $k_{UV}\sim aH$. Away from this approximation, our result will receive an additional term which depends on the UV cutoff.  Our graviton mass has nothing to do with the UV physics, thus it isn't our main interest and we are not going to discuss the UV divergence issue in this paper.


{\bf Conclusion and Discussion}
In this short notes, we consider a spatial condensation scenario, which background solution spontaneously breaks the spatial diffeomorphism.   In the unitary gauge, massless graviton eats the Goldstone excitations of spatial condensation, gets weight and becomes a massive graviton. Our massive graviton is a multiplet particle, its 5 polarizations have exactly the same dispersion relation, with a mass gap on the spectrum. 

We then apply our massive gravity theory  to inflation, and find graviton mass removes the IR divergence of inflationary loop diagram.  In addtion to the virtue of IR safe, we would expect our model has some other interesting features. The primordial tensor mode may receive a modification due to the graviton mass,  and we expect to find some interesting feature on the B mode polarization of CMB\cite{solitonMG}. 

Although we only checked the stability of our theory at FRW background, we expect it has the universal healthy nature since our theory is nothing but Einstein-Hilbert action and 3 canonical free scalars. More generally, taking the SO(3) symmetry of scalars' configuration as our building principle, we can write down a most general action with non-derivative graviton potential terms as 
\begin{eqnarray}
S=M_p^2\int\sqrt{-g}\left[\frac{\mathcal{R}}{2}-m^2\mathcal{U}\left(g^{\mu\nu},f_{\mu\nu}\right)\right].
\end{eqnarray}
where $f_{\mu\nu}\equiv\partial_{\mu}\phi^a\partial_{\nu}\phi^b\delta_{ab}$ and $\mathcal{U}\left(g^{\mu\nu},f_{\mu\nu}\right)$ is a general function of $g^{\mu\nu}$ and $f_{\mu\nu}$. Besides the non-derivative potential terms, we are also able to introduce the derivative coupling terms, e.g. the Horndeski term $G^{\mu\nu}f_{\mu\nu}$, where $G^{\mu\nu}$ is the Einstein tensor. The stability of such theory is checked in the ref. \cite{Lin:2013aha}.

Last but not least, it is worth to notice that the idea of spatial condensation is actually not new. 
Such kind of scalar fields configuration generally appears at soliton physics. Let's take the simplest global monopole as an example. Considering a monopole as big as a universe, e.g. a topological inflation\cite{Linde:1994hy}\cite{Vilenkin:1994pv},  we have 3 canonical scalar fields with nontrivial background VEV at 3 spatial direction inside of monople. According to the analysis in this paper, we expect that graviton inside of monopole appears to be massive. The relevant study will be covered in our future work \cite{solitonMG}.




~~~
\begin{acknowledgments}
{\bf acknowledgments}
The author would like to thank A. Emir Gumrukcuoglu, G. Gabadadze, K. Hinterbichler,  R. Kimura, L. Labun, S. Mukohyama, R. Saito, M. Sasaki, G. Shiu, N. Tanahashi, T. Tanaka, H. Tye,  Y. Wang, W. Xue for the useful discussion. The author also would like to thank the hospitality of Yukawa institute, since the idea of this paper was spontaneously generated on the author's way back from Yukawa institute, after two days' short visiting. This work is supported by the World Premier International Research Center Initiative (WPI Initiative), MEXT, Japan. 
\end{acknowledgments}

\end{document}